\documentclass{ws_ijmpa}
\newcommand{\rp}{\mbox{\boldmath $p$}}

\begin{document}

\markboth{M. B. Gay Ducati, V. P. Gon\c calves and L. F. Mackedanz}
{Collisional energy loss with running $\alpha_s$}

%
\catchline{}{}{}{}{}
%

\title{\bf QCD COLLISIONAL ENERGY LOSS IN AN INCREASINGLY INTERACTING QUARK GLUON PLASMA}

\author{M. B. GAY DUCATI}

\address{Instituto de F\'{\i}sica, Universidade Federal do Rio Grande do Sul\\
Caixa Postal 15051, CEP 91501-970, Porto Alegre, RS, Brazil\\
beatriz.gay@ufrgs.br}

\author{V. P. GON\c CALVES}

\address{Instituto de F\'{\i}sica e Matem\'atica,  Universidade Federal de Pelotas\\
Caixa Postal 354, CEP 96010-090, Pelotas, RS, Brazil\\
barros@ufpel.edu.br}

\author{L. F. MACKEDANZ $^{a,b}$}

\address{$^a$ Centro de Ci\^encias Exatas e Tecnol\'ogicas, Universidade Federal de Pelotas\\
Campus Ca\c capava do Sul\\
Rua Rui Vieira Machado s/$n^{o}$, CEP 96570-000, Ca\c capava do Sul, RS, Brazil\\
$^b$ Instituto de F\'{\i}sica, Universidade Federal do Rio Grande do Sul\\
Caixa Postal 15051, CEP 91501-970, Porto Alegre, RS, Brazil\\
luiz.mackedanz@ufrgs.br}

\maketitle


\begin{abstract}
The discovery of the  jet quenching in central Au + Au collisions at the Relativistic Heavy-ion Collider (RHIC) at Brookhaven National Laboratory has provided clear evidence for the formation
of strongly interacting dense matter. It has been predicted to occur due to the energy loss of high energy partons that propagate through the quark gluon plasma. In this paper we investigate the
dependence of the parton energy loss due to elastic scatterings in a parton plasma on the value of the strong coupling and its running with the evolution of the system. We analyze different
prescriptions for the QCD coupling and calculate the energy and length dependence of the fractional energy loss. Moreover, the partonic quenching factor for light and heavy quarks is estimated. We found that the predicted  enhancement of the heavy to light hadrons ($D/\pi$) ratio  is strongly dependent on the running of the QCD coupling constant.

\end{abstract}

\ccode{PACS numbers: 12.38.Mh; 13.85.N; 25.75.-q}

\section{Introduction}

Relativistic heavy ion collisions provide an  opportunity to study
the QCD properties at energy  densities about thirty times higher
than the density of the atomic nuclei \cite{adams,rafelski,wong}. In
these extreme conditions a deconfined state of quarks and gluons,
the Quark Gluon Plasma (QGP), is expected to be formed in the early
stage of the collision. Recently, the discovery of jet quenching in
central Au + Au collisions at the Relativistic Heavy-ion Collider
(RHIC) at Brookhaven National Laboratory has provided clear evidence
for the formation of strongly interacting dense matter. Detailed
analysis indicate that the suppression of the single hadron spectra
at high $p_T$, the disappearance of back-to-back correlation of high
$p_T$ hadrons and the azimuthal anisotropy of high $p_T$ hadron
spectra in noncentral collisions  are caused by parton energy loss
(For a recent review see, e.g, Ref. \cite{wang_jacobs}). Basically,
the high parton density produced in heavy ion collisions could
induce a large amount of energy loss while hard partons produced in
the initial stage of the collision propagate through the fireball,
due to the interactions of the hard partons with the medium.

The total energy loss of a particle in a medium  can be decomposed
into a collisional and a radiative contribution. While the first one
originates from the energy transfer to the medium particles, the
latter one is caused by radiation from the fast particle. At large
energies one expects that radiative energy loss becomes much larger
than the collisional one, as in the electromagnetic case. However,
at lower energies these two processes can contribute equally, with
the collisional one being the dominant for small values of the
parton energy. Currently, an open question is to quantify the
contribution of each process in the RHIC kinematical region. In
particular, in the last few years the understanding of parton energy
loss by gluon bremsstrahlung has been extensively developed (For
recent reviews see, e.g., Refs.
\cite{baier_schiff,wang_QGP3,kovner_QGP3,hard_probes_cern}). Recent
works have analyzed the energy \cite{wang_prc_energy,gyulassy_adil},
color charge and mass dependence of radiative parton energy loss
\cite{armesto_dainese,wang2zhang,gyulassy_magda}, as well as its
non-Abelian feature \cite{wang_nonabelian}. A basic characteristic
from these works is the assumption  that the radiative energy loss
dominates disregarding the collisional one. However, recent studies
in charm quark thermalization \cite{rapp,guy_moore}, quenching of
hadron spectra \cite{mustafa_thoma,mustafa} and the elastic parton
energy loss including all $2 \rightarrow 2$ processes
\cite{stopping}  indicate that in the RHIC kinematical region it is
far from clear that radiative energy loss dominates over collisional
energy loss.

Another important  aspect is that in  general the approaches for
radiative and collisional energy loss start from the assumption that
the properties of the medium and its interactions with the energetic
parton projectile do not change with time (For a discussion of the
medium evolution see, e.g. Refs. \cite{vitev,salgado}). However, in
nucleus - nucleus collisions at collider energies, the produced hard
partons propagate through a rapidly expanding medium. The density of
scattering centers is expected to reach a maximum value at the
plasma formation time, $\tau_0$, and then decrease with time $\tau$
rapidly due to the strong longitudinal expansion. We begin following
the progress of the quark at $\tau = \tau_0 $, where it begins a
process of scattering and diffusion in the plasma which causes a
loss of the initial momentum and relaxation towards the thermal
velocity. The scenario which we assume in this paper for the
equilibration has been proposed some years ago \cite{wongAlpha} (See
also Ref. \cite{nayak_mclerran}) and consider that the interactions
in the plasma of quarks and gluons get stronger with the time
because the average parton energy drops due to the expansion of the
system. The basic idea comes from the feature that  the QGP is not a
static medium, but it is cooling while partons propagate through.
Consequently, the scale of the coupling changes with the evolution
of the system, and this feature motivates the calculation of the
observables assuming the running coupling. In this paper we will
assume that the temperature is the dominant scale and consequently
will control the running of the QCD coupling. The standard procedure
is to take the solution for the running coupling from the
renormalization group equation, which to lowest (1 - loop) order is
given by:
\begin{equation}
\alpha_s (\overline{\mu}) = \frac{12 \pi}{(33-2 n_f)\ln
\frac{\overline{\mu}^2}{ \Lambda_{\overline{MS}}^2}} \,\,,
\label{alphas}
\end{equation}
and put the  renormalization point $\overline{\mu}$ proportional to
the first Matsubara frequency, $\overline{\mu} \propto  2 \pi T$,
which for massless quarks is the only dimensionfull quantity
inherent to the theory. Moreover, the strong coupling $\alpha_s$
also depends on the scale parameter of the modified minimal
subtraction scheme $\Lambda_{QCD}$. However, it is not evident that
temperature is the relevant scale at the energies currently
accessible in heavy ion physics,  where the temperatures reached are
only moderately larger than the phase transition temperature. As the
temperature drops over the lifetime of the QGP, $\alpha_s$ should
also vary during the equilibration and the evolution of the plasma.
In order to simplify our considerations we model the space-time
evolution of the quark-gluon plasma by the Bjorken scenario with
boost invariant longitudinal expansion and conserved entropy per
rapidity unit \cite{bjorken}. This implies that $T = T_0
(\frac{\tau_0}{\tau})^{\frac{1}{3}}$, where $T_0$ is the initial
temperature at the initial time $\tau_0$. We neglect the transverse
expansion of the system. Furthermore, we will consider in what
follows two prescriptions for the temperature dependence of the
running coupling constant. First we will assume
\begin{eqnarray}
 \alpha_s (T) = \frac{6 \pi}{(33-2 n_f)\ln[(19
T_c/\Lambda_{\overline{MS}})(T/T_c)]}, \label{alpha1}
\end{eqnarray}
where $T_c/\Lambda_{\overline{MS}} = 1.78\pm0.03$.  This
prescription, which we denote by  thermal [$\alpha_s^{(th)}$], has
been used in, e.g., Refs. \cite{gunion_vogt,vogt_levai}.
Furthermore, we also consider that $\alpha_s$ can be given by
\begin{eqnarray}
\alpha_s(T) = \frac{2.095}{\frac{11}{2 \pi} \ln
\left(\frac{Q}{\Lambda_{\overline{MS}}}\right) + \frac{51}{22 \pi}
\ln\left[2 \ln
\left(\frac{Q}{\Lambda_{\overline{MS}}}\right)\right]},
\label{alpha2}
\end{eqnarray}
with $Q = 2 \pi T$.  This parametrization of the strong coupling has
been obtained from recent results in the lattice \cite{kaczamarek}.
We denote this prescription by $\alpha_s^{(\mathrm{lat})}$.

\begin{figure}[pb]
\vspace*{1.0cm}
\centerline{\psfig{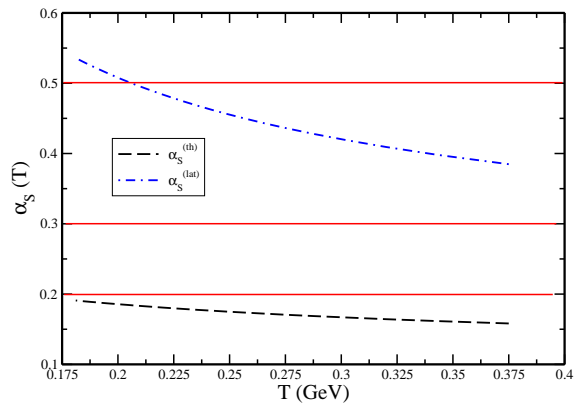}}
\vspace*{8pt} \caption{Temperature dependence of the QCD coupling
considering two different prescriptions. See text. } \label{fig0}
\end{figure}

In Fig. \ref{fig0} we present the temperature  dependence of
$\alpha_s$ predicted by these two prescriptions. We use  $T_0 = 375$
MeV and $t_0 = 0.33$ fm which are reliable values for RHIC energies
\cite{rapp}.  The horizontal lines characterize  the constant values
of the QCD coupling: (a)  $\alpha_s = 0.3$, as in previous
calculations \cite{mustafa},  (b) $\alpha_s = 0.2$, which is a
typical value for the coupling constant and (c) $\alpha_s = 0.5$ for
comparison. We can see that there is a large difference in
normalization and shape between the two prescriptions. Comparing the
results, the lattice motivated prescription given in Eq.
(\ref{alpha2}) [$\alpha_s^{(lat)}$ in the figure] provides the
largest $\alpha_s$ value at small temperatures. The fixed values
$\alpha_s = 0.2$ and 0.3  lie between these two prescriptions. On
the other hand, a fixed value of $\alpha_s = 0.5$ is larger than the
lattice prescription at large temperatures. It is important to
emphasize that if the prescriptions proposed in Refs.
\cite{wongAlpha,rafelski_tounsi} are used we obtain a behavior for
the running coupling similar to  $\alpha_s^{(\mathrm{lat})}$.

As the temporal development of the coupling constant modifies  the
behavior of  several QGP signatures
\cite{wongAlpha,nayak_mclerran,rafelski_tounsi,wong_spectra,denterria},
we can expect a similar effect in the estimates of the energy loss
of a parton propagating in a QGP. Our main goal in this paper is to
estimate the influence of the running coupling constant in the
collisional energy loss, presenting a  reanalyzes of the studies
from Refs. \cite{mustafa_thoma,mustafa,svetitsky,dipali}, and an
estimate of the quenching factor $Q(p_\perp)$ for light and heavy
quarks (See discussion in Sect. \ref{elastic}). We postpone for a
future publication the study of this effect in the radiative energy
loss (For a short discussion see Ref. \cite{zakharov_jetp_lett}).

A comment is in order here. In our  studies  we  consider that the
fractional energy loss is given in terms of the Bjorken formula
\cite{bjorken_1982} as  generalized in Refs.
\cite{thoma_ela,braaten_thoma}. Recently, Peshier \cite{peshier} has
advocated that the collisional energy loss is an observable for
which loop corrections to the tree level approximation are
essential. In particular,  if the running of the coupling constant
is taken into account the collisional energy loss becomes
independent of the jet energy  in the high energy limit ($E \gg T$).
It implies  a distinct temperature dependence for the elastic energy
loss when compared to the Bjorken formula and a larger value for the
mean energy loss per length. If the formalism considered in this
paper is reanalyzed assuming as input the Peshier's prediction,  we
can expect a larger  contribution of the elastic energy loss.
Consequently, the   results presented here can be considered as a
lower bound for the contribution of the elastic energy loss for the
parton quenching.

This paper is organized as follow. In next section we  define the
partonic quenching factor $Q(p_{\perp})$ and present its relation
with the energy loss $\Delta E$. Furthermore, the Fokker-Planck
equation is derived and its solution is obtained. In Sect.
\ref{results} we present our predictions for the time evolution of
the drag coefficient for the propagation of light and heavy quarks
considering different prescriptions for the QCD coupling, as well as
our results for the energy and length dependence of the fractional
energy loss. Moreover, our predictions for the quenching factor for
light and heavy quarks are also presented. Finally, in Sect.
\ref{conc} we present some remarks and our main conclusions.

\section{Parton Propagation in a QGP and the Partonic Quenching Factor}
\label{elastic}

Lets start our discussion considering the hadron production in
nucleus-nucleus collisions. Due to the interactions of the  produced
parton with the medium, it loss an additional energy fraction
$\Delta E$ while escaping the collision. Consequently, the inclusive
transverse momentum spectra of the particles produced in nucleus -
nucleus collisions will be modified with respect to hadron - hadron
collisions. In general, the hadron formation is described in terms
of parton recombination and/or by the fragmentation of the energetic
partons. In particular, it is expected that for $p_\perp > 5$ GeV
the hadrons are dominantly produced by fragmentation. In what
follows we  assume that fragmentation is the dominant process of
hadron formation and that it occurs after the parton has left the
comoving medium. It allows to analyze the energy loss effects
directly in the $p_\perp$ spectrum of the scattered partons. In
Refs. \cite{bdms,mueller} the effect of the radiative parton energy
loss in the  $p_\perp$ distribution has been estimated. Here we
extend this approach for collisional energy loss. Following Refs.
\cite{bdms,mueller} we assume that the $p_\perp$ spectrum is given
by:
\begin{eqnarray}
\frac{dN^{\rm{med}}}{d^2p_\perp} &=& \int d\epsilon \, D(\epsilon)
\, \frac{dN^{\rm{vac}}(p_\perp+\epsilon)}{d^2p_\perp} \equiv
Q(p_\perp) \frac{dN^{\rm{vac}}(p_\perp)}{d^2p_\perp}, \label{rate}
\end{eqnarray}
where  $\frac{dN^{\rm{vac}}(p_\perp+\epsilon)}{d^2p_\perp}$ is the
transverse momentum distribution in elementary parton-parton
collisions, evaluated at a shifted value $p_\perp+\epsilon$, and
$D(\epsilon )$ is the probability distribution in the energy
$\epsilon$ lost by the partons in the medium through collisions.
Moreover, $Q(p_\bot)$ is the medium dependent quenching factor. In
Ref. \cite{bdms} the authors have demonstrated that in a realistic
calculation of the quenching, the knowledge of the full probability
distribution is actually required. However, as our goal is to study
the effect of the running of coupling constant in the estimates of
the quenching factor, we will assume, following Ref. \cite{mueller},
that the quenching can be modeled in terms of the mean energy loss.
Therefore, in this work we will calculate the partonic quenching
factor assuming that it can be approximated by \cite{mueller}
\begin{eqnarray}
Q(p_\bot) =
\frac{dN^{\rm{med}}}{d^2p_\perp}/\frac{dN^{\rm{vac}}}{d^2p_\perp}
\end{eqnarray}
where
\begin{equation}
\frac{dN^{\rm{med}}}{d^2p_\perp}= \, \frac{1}{2\pi^2 R^2} \
\int_0^{2\pi} \, d\phi \, \int_0^R \, d^2r \, \,
\frac{dN^{\rm{vac}} (p_\perp+ \Delta E)}{d^2 p_\perp}\, \, \,,
 \label{supp}
\end{equation}
$R$ is the nuclear radius, $\phi$ is the angle between the velocity
and the radius vector of the parton [See Eq. (\ref{trad})] and
$\Delta E$ is the total energy loss by partons in the medium.

In order to calculate $\Delta E$ we  will use the approach proposed
by Svetitsky in Ref. \cite{svetitsky}, which considers the Brownian
motion of a parton in a thermal bath, governed by the Fokker-Planck
equation. This approach has been used to estimate the diffusion of
charm quarks in a quark - gluon plasma \cite{svetitsky,dipali} (See
also Ref. \cite{rafelski_prl}) and its equilibration
\cite{alam_prl,alam_npa}, as well as to calculate the quenching of
light \cite{mustafa_thoma} and heavy quarks
\cite{rapp,guy_moore,mustafa}. In this approach one starts from  the
Boltzmann equation for the distribution function $f (x,p)$ and
assume that there is no external force acting on the quark and that
the phase space distribution $f$ does not depend on the position of
the quark. Furthermore, assuming the scattering process to be
dominated by small momentum transfers, one arrives at a
Fokker-Planck equation describing the evolution of $f$ in the
momentum space. We then investigate the time evolution of the
Fokker-Planck equation in a thermally evolving QGP. The main
simplifying assumption in the approach  is the momentum-independence
of the drag and diffusion coefficients. In what follows we present
the main formulae of this approach. However, we do not repeat here
the details of the calculations and refer the reader to the original
works \cite{svetitsky,dipali,mustafa_thoma,mustafa}.

The Boltzmann equation in the  relativistically covariant form can
be written as \cite{bales}
\begin{equation}
p^\mu\partial_\mu f(x,p) = C\{ f \} \, , \label{boltz}
\end{equation}
where $p^\mu = (E_{\rp},\rp)$ is  the four-momentum of the test
quark and $f$ is its phase-space density and $C\{ f\}$ is the
collision term. Following Refs.
\cite{svetitsky,dipali,mustafa_thoma,mustafa} we assume that: (a)
the hydrodynamical evolution can be described by the Bjorken
scenario, which implies that it is valid to assume that the plasma
is uniform and consequently the phase space density  of the quark is
independent of $\vec{x}$, and  (b) the collision term is given by
the elastic collisions of the test quark with other quarks,
antiquarks and gluons in the system. Using  the  Landau
approximation and restringing the analyzes to the one-dimensional
problem one obtain \cite{svetitsky,dipali,mustafa_thoma,mustafa}
\begin{equation}
\frac{\partial f}{\partial t} = \frac{\partial}{\partial p} \left[
{\mathcal T}_1(p) f \right] + \frac{\partial^2}{\partial p^2}
\left[ {\mathcal T}_2(p) f \right],
 \label{fp0}
\end{equation}
which is the Landau kinetic equation,  with the transport coefficients given by
\begin{eqnarray}
{\mathcal T}_1(p) &=& \int
dk \, w(p,k)\, k = \frac {\langle \delta p \rangle}
{\delta t} = \langle F \rangle \, , \nonumber \\
{\mathcal T}_2(p) &=&\frac{1}{2} \int d^3k\, w(p,k)\, k^2 = \frac
{\langle (\delta p )^2\rangle} {\delta t} \, \, , \label{coeff}
\end{eqnarray}
where $ w(p,k)$ sums the  rate of collisions of a test particle with
the partons of the medium  and $\langle F \rangle$ is the average
force acting on the test particle. If we consider that the
background heat bath is constituted of a large amount of  weakly
coupled particles in thermal equilibrium at a temperature T, with
some non-thermal but homogeneously distributed particles due to the
fluctuations, the problem can be simplified \cite{bales} assuming
that the equilibrium of the bath will not be disturbed by the
presence of these few non-thermal particles. Due to its small
number, one can also assume that they will not interact among
themselves, only with particles of the thermal bath. Consequently,
one can replace the distribution functions of the collision partners
of the test particle by their Fermi-Dirac or Bose-Einstein
distributions and Eq. (\ref{fp0}) reduces to the Fokker-Planck
equation.

The transport coefficients ${\mathcal T}_1$ and  ${\mathcal T}_2$
can be expressed in terms of the drag coefficient of quark
${\mathcal A}$, which is almost independent of momentum $p$ as shown
in Refs. \cite{svetitsky,dipali}. It allows to write the
Fokker-Planck equation as follows
\cite{svetitsky,dipali,mustafa_thoma,mustafa}
\begin{equation}
\frac{\partial f}{\partial t} = {\mathcal A} \frac{\partial
}{\partial p} (p f) + {\mathcal D}_F\frac{\partial^2 f}{\partial
p^2} \, \, , \label{landau}
\end{equation}
where ${\mathcal D}_F$ is the diffusion  coefficient, which is given
by  $ {\mathcal D}_F = {\mathcal A} T^2 $ if we assume that the
momentum $p$ can be approximated by the temperature $T$ of the
system and the coupling between the Brownian particle and the bath
is weak \cite{bales}. The Eq. (\ref{landau}) describes the evolution
of the momentum distribution of a test particle undergoing Brownian
motion. The solution from Eq. (\ref{landau}) has been obtained in
Refs. \cite{mustafa_thoma,mustafa}, assuming as   boundary condition
$ f(p,t)\, \, \, \stackrel{t\rightarrow t_0}{\longrightarrow} \, \,
\delta(p-p_0)$ and using the  method of characteristics
\cite{william}. It is given by \cite{mustafa_thoma,mustafa}
\begin{eqnarray}
f(p,L) &=& \frac{1}{\sqrt{\pi\, {\mathcal W}(L)}} \, \exp \left [
- \frac{\left (p-p_0\, e^{-\int^L_0{\mathcal A}(t') \, dt'} \right
)^2} {{\mathcal W}(L)} \right ] \, \, , \label{solfin}
\end{eqnarray}
where ${\mathcal W}(L)$ is given by
\begin {equation}
{\mathcal W}(L) = \left ({4\int_0^L {\mathcal D}_F(t') \exp \left [
2 \int^{t'}_0 {\mathcal A}(t'')\, dt'' \right ]\, dt'}\right ) \left
[{\exp \left (-2 \int_0^L {\mathcal A}(t')\, dt'\right )} \right ]
\, \, , \label{gauswid1}
\end{equation}
 and is the probability distribution in  momentum space. In the previous equation the
length of the expanding plasma is assumed as being  the maximum time
limit.

From Eq. (\ref{solfin})  we can estimate the mean energy of the
parton due to elastic collisions after traversing a distance $L$. It
is given by
\begin{equation}
\langle \, E \, \rangle \, = \, \int_0^\infty \, E \, f(p,L) \, dp
\, \, . \label{meane}
\end{equation}
Then the average energy loss due to elastic collisions in the medium
will be given by
\begin{eqnarray}
\Delta E\,   &=& \, E_0\, -\, \langle \, E \, \rangle  \, \, \, ,
\label{avgelos}
\end{eqnarray}
where  $E=m_\perp \, = \sqrt{p_\perp^2 \, + M^2 }$ at the central
rapidity region, $y=0$. Consequently, in order to estimate the
average energy loss it is necessary to calculate Eq. (\ref{solfin})
in terms of the drag coefficient of a quark.  Following Refs.
\cite{svetitsky,dipali}, we approximate the drag coefficient by its
average value,
\begin{equation}
\langle {\mathcal A}(p,t) \rangle \, = \, {\mathcal A}(t) \, = \,
\left \langle -\frac{1}{p}\, \, \frac{dE}{dL}
 \right \rangle \, \, ,
\label{average}
\end{equation}
which is   directly dependent of the energy loss rate $dE/dL$. The
above approximation is  reasonable  up to moderate momentum values
($p \leq 15$ GeV) \cite{mustafa}.

The energy  loss rate in the QGP due to elastic collisions with
high-momentum transfer have been originally estimated by Bjorken
\cite{bjorken_1982} and recalculated in Refs.
\cite{mrom,thoma_ela,braaten_thoma,thoma_epjc} taking into account
the loss with low-momentum transfer dominated by the interactions
with plasma collective modes in the hard thermal loop approximation
\cite{braaten_pisa}. In particular, in Ref. \cite{braaten_thoma} the
authors have estimated the energy loss for heavy quarks and in Ref.
\cite{thoma_ela} for light partons. For heavy  quarks and in the
domain $E<<M^2/T$, it reads
\begin{equation}
-\, \frac{dE}{dL}\, = \, \frac{8\pi\alpha_s^2T^2}{3} \left
(1+\frac{n_f}{6}\right ) \left [ \frac{1}{v} - \frac{1-v^2}{2v^2}
\ln \left (\frac{1+v}{1-v}\right )\right ] \, \ln\left [
2^{\frac{n_f}{6+n_f}} B(v) \frac{ET}{m_g M}\right ]  \,
\label{rate1}
\end{equation}
whereas for $E>>M^2/T$, it is
\begin{eqnarray}
-\, \frac{dE}{dL}\, &=& \, \frac{8\pi\alpha_s^2T^2}{3} \left
(1+\frac{n_f}{6}\right ) \, \ln\left [ 2^{\frac{n_f}{2(6+n_f)}}
0.92 \frac{\sqrt{ET}}{m_g }\right ]  \, \label{rate2}
\end{eqnarray}
where $n_f$ is the number of quark flavors,  $\alpha_s$ is the
strong coupling constant, $m_g= \sqrt {(1+n_f/6)g^2 T^2/3}$ is the
thermal gluon mass, $E$ is the energy and $M$ is the mass of the
quark. $B(v)$  is a smooth velocity function, which can be taken
approximately as  $0.7$ \cite{braaten_thoma}. For light quarks we
use the expression (\ref{rate2}) and set $M = 0$ in the
calculations. Inclusion of the diagrams other than the $t$ channel
increase the energy loss rate for light quarks by a factor 2
\cite{stopping}.

\section{Results and Discussion}
\label{results}

\begin{figure}[htbp]
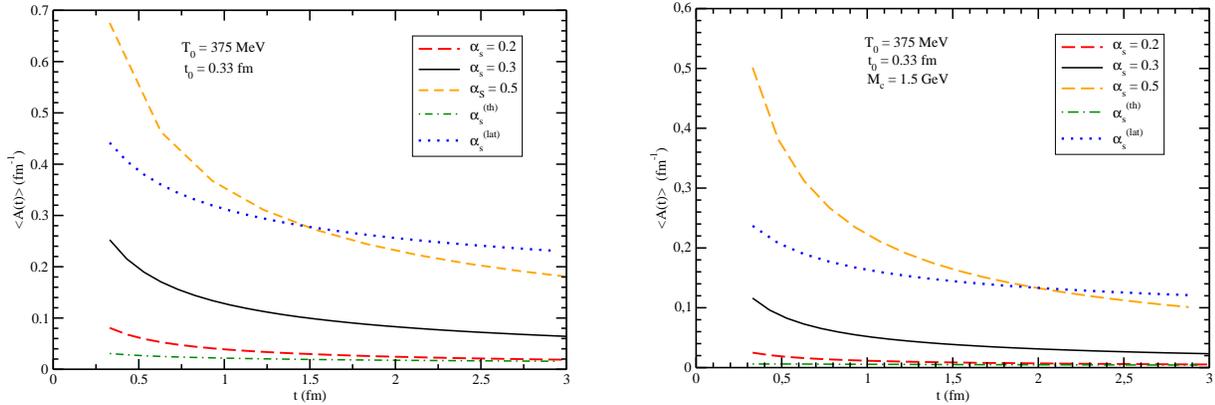

\vspace*{0.5cm} \centerline{\begin{tabular}{lcr}
\hspace*{-0.2cm}\psfig{file=dragcoeflight2.eps,width=75mm} & &
\hspace*{0.2cm}\psfig{file=dragcoefcharm2.eps,width=75mm}
\end{tabular}}
\vspace*{8pt}
\caption{Time evolution of drag coefficient for light (left panel) and
heavy (charm) quarks (right panel).}
\label{fig1}
\end{figure}

At the energies (temperatures) which we are interested  in this
paper the drag coefficient  ${\mathcal A}$ for partons propagating
in a plasma can be calculated using Eq. (\ref{average}) and the
expressions for elastic energy loss given by Eqs. (\ref{rate1}) and
(\ref{rate2}). The average over the momentum is made using the
Boltzmann distribution. As we assume the Bjorken scenario for the
hydrodynamics evolution \cite{bjorken}, the time dependence of the
temperature is given by $T(t)=t_0^{1/3} T_0/t^{1/3}$, where $t_0$
and $T_0$ are, respectively, the initial time and temperature at
which the background of the partonic system has attained local
kinetic equilibrium. The time dependence from the drag coefficient
is directly associated with this evolution for the temperature,
which decreases with time as the system expands. We assume as
maximum time limit for the evolution the length of the plasma $L$.
These approximations has been considered in Refs.
\cite{mustafa_thoma,mustafa} which we would like compare our
results.  Moreover, we assume $T_0 = 375$ MeV and $t_0 = 0.33$ fm
for RHIC energies  as in Ref. \cite{rapp}. The results for the time
dependence of the drag coefficient ${\mathcal A}$ are show in Fig.
\ref{fig1} for light and heavy (charm) quarks. We present a
comparison among the results for fixed $\alpha_s$  and running
coupling from lattice and thermal QCD prescriptions. One can see
that the drag coefficient is very sensitive to the modifications in
the prescription used for $\alpha_s$. We have that when the lattice
one is considered (dotted line), the coefficient becomes larger than
earlier results for fixed $\alpha_s = 0.3$ (solid line)
 in all the evolution of the fireball. In comparison to the $\alpha_s
= 0.5$ case (dashed line), the lattice prescription implies a
smaller value of ${\cal{A}}(t)$ for small values of $t$. At large
times, $\alpha_s^{(lat)}$ becomes larger than 0.5 due to  the
cooling of the system, which implies a larger value of
${\cal{A}}(t)$.  On the other hand, when the thermal QCD
prescription is considered (dot-dashed line), the drag coefficient
is smaller than the previous result, being closer to the prediction
obtained assuming  $\alpha_s = 0.2$ (long-dashed line).

\begin{figure}[ht]
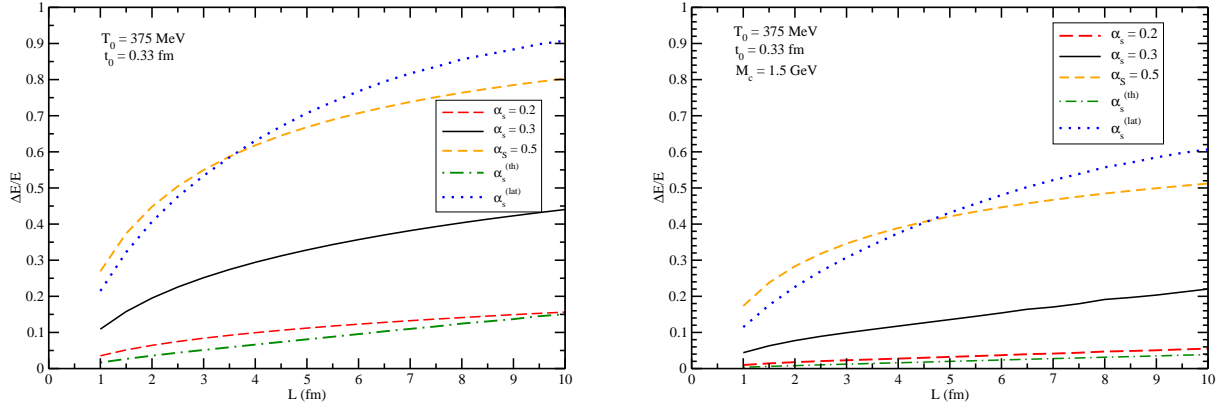

\centerline{\begin{tabular}{lcr}
\hspace*{-0.2cm}\psfig{file=Ldependencelight2.eps,width=75mm} &  &
\hspace*{0.2cm}\psfig{file=Ldependencecharm2.eps,width=75mm}
\end{tabular}}
\vspace*{8pt}
\caption{Fractional energy loss as a function of the distance traveled by
the light (left panel) and heavy quark (right panel).}
\label{fig2}
\end{figure}

In Fig \ref{fig2} we  present  the results for the fractional energy
loss, as a function of the distance traveled by the partons (i.e,
the time evolution of the plasma) for light and heavy quarks.
Following the results for the drag coefficient, shown in Fig.
\ref{fig1} and discussed above, we have that  $\alpha_s = 0.5$
implies the larger amount of energy loss at small values of $L$. On
the other hand, the lattice QCD $\alpha_s$ prescription implies a
large energy loss for large $L$. The thermal QCD $\alpha_s$
prescription provides the smallest one. Moreover, as expected, the
fractional energy loss for heavy quarks is smaller than for light
quarks.

\begin{figure}[ht]
\vspace*{0.5cm} \centerline{\psfig{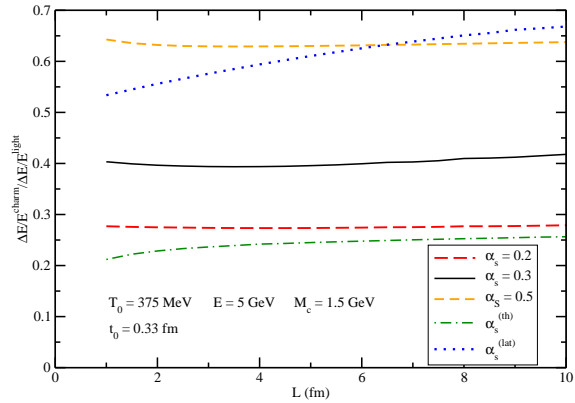}}
\vspace*{8pt} \caption{Heavy-to-light ratio of the fractional energy
loss.} \label{fig3}
\end{figure}

\begin{figure}[pb]
\vspace*{0.5cm}
\centerline{\psfig{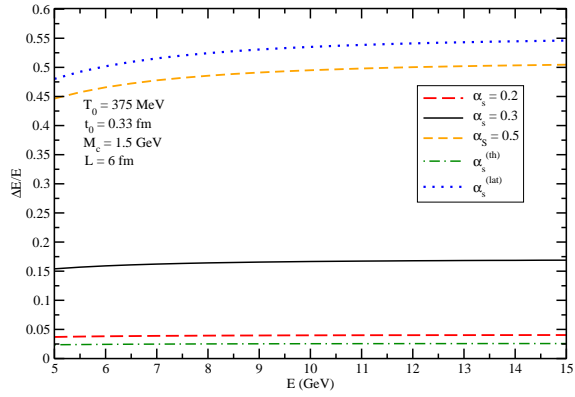}}
\vspace*{8pt} \caption{Heavy quark fractional energy loss as a
function of the energy of incident quark.} \label{fig4}
\end{figure}

In Fig. \ref{fig3} we present  the ratio between the fractional
energy loss for heavy  and light quarks. While for fixed $\alpha_s$
the ratio is almost constant in the range considered, for running
$\alpha_s$ with the lattice prescription, the ratio is monotonously
increasing, and the value is greater than $0.5$ in all range
considered; on the other hand, using  the thermal prescription, the
ratio is strongly suppressed when compared with the result for fixed
$\alpha_s$. This feature suggests that the heavy quarks lose less
than $20 \%$ of the energy lost by light quarks in its path through
the fireball. The predictions obtained using $\alpha_s = 0.5$ are
similar to the lattice one.

In Fig. \ref{fig4}  we present the  results for the fractional heavy
quark energy loss as a function of the incident quark energy. One
have that the predictions are almost energy independent in the
energy range $E \sim 5-15$ GeV for all prescriptions analyzed. This
feature is due to the momentum independence of the drag coefficient,
as discussed earlier \cite{svetitsky,dipali}. Furthermore, the
magnitude of the energy loss is strongly dependent  on the
prescription used in the calculation.

\begin{figure}[pb]
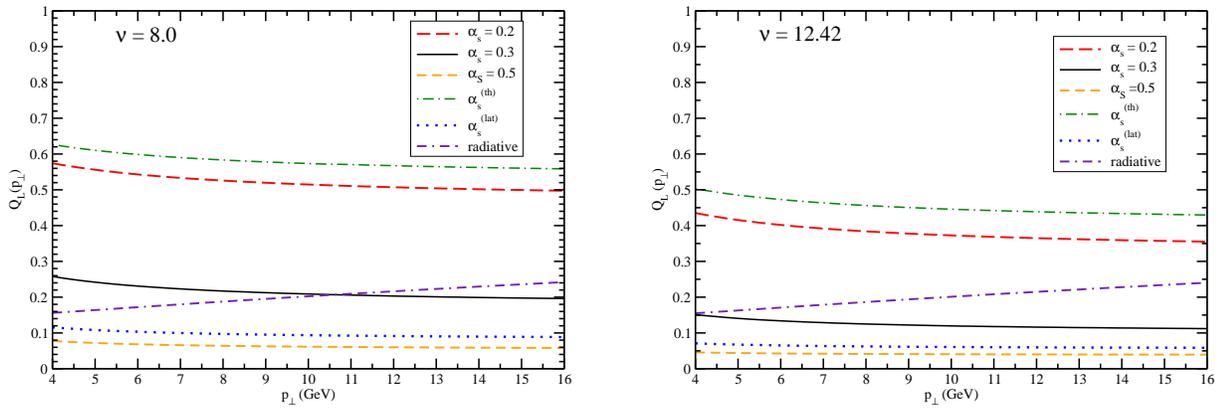

\vspace*{0.5cm} \centerline{\begin{tabular}{lcr}
\hspace*{-0.2cm}\psfig{file=quenchinglight8_2.eps,width=75mm} &  &
\hspace*{0.2cm}\psfig{file=quenchinglight12_2.eps,width=75mm}
\end{tabular}}
\vspace*{8pt} \caption{Quenching factor for light quark
$p_\perp$-spectrum. An estimative of radiative energy loss is
presented for comparison. The left panel presents results with $\nu
= 8.0$ in the vacuum spectrum parametrization, while results with
$\nu = 12.42$ are presented in the right panel.} \label{fig5}
\end{figure}

In order to compute  the $p_\perp$-spectra,  we assume that the
geometry of the heavy ion collision is described by a cylinder of
radius $R$ and the parton moves in the transverse plane in the local
rest frame. Consequently, if we consider a parton created at a point
$\vec{\mathbf r}$ with an angle $\phi$ in the transverse direction
it will travel a distance \cite{mueller}
\begin{equation}
L(\phi)= (R^2\, - \, r^2 \, \sin^2\phi \,)^{1/2}\, - \, r \, \cos \phi \, \,\, ,
\label{trad}
\end{equation}
where $\cos \phi\, = \,{\hat {\vec {\mathbf v}}} \, \cdot \,
{\hat{\vec {\mathbf r}}}\, \, $; ${\vec {\mathbf v}}$ is  the
velocity of the parton and $r\, = \, |{\vec {\mathbf r}}|$.
Furthermore, we  assume the following parametrization of the
$p_\perp$ distribution
\begin{equation}
\frac{dN^{\rm{vac}}_L}{ d^2p_\perp}\, = \, A \left (\frac{1}{p_0+p_\perp}
\right )^{\nu} \, \, \, ,
\label{lparam}
\end{equation}
where two sets of parameters are  available in the literature: $\nu
= 8.0$ and $p_0=1.75$ GeV \cite{mueller} and $\nu = 12.42$ and
$p_0=1.71$ GeV \cite{dokskhar}. The results for the quenching factor
for light quarks are  shown in the Fig. \ref{fig5} for both sets of
parameters. For  comparison, we present the  estimate of the
quenching due to the radiative energy loss, following the
parametrization proposed in Ref. \cite{mueller}. Due to the smaller
drag coefficient, the thermal QCD $\alpha_s$ prescription gives a
high quenching factor, so the spectrum is less modified by
collisional energy loss than in the case of fixed $\alpha_s$. At
$\alpha_s = 0.3$ and $\nu = 8.0$ we have that in the high $p_\perp$
region, elastic and radiative energy loss are of the same order of
magnitude. On the other hand, the lattice QCD $\alpha_s$
prescription implies a very large modification of the spectrum,
similar to the predictions obtained assuming $\alpha_s = 0.5$.
Finally, if the system presents a lower value of $\alpha_s$ than
considered in earlier calculations, the gluon bremsstrahlung becomes
the dominant mechanism of energy loss again.

\begin{figure}[pb]
\vspace*{0.5cm}
\centerline{\psfig{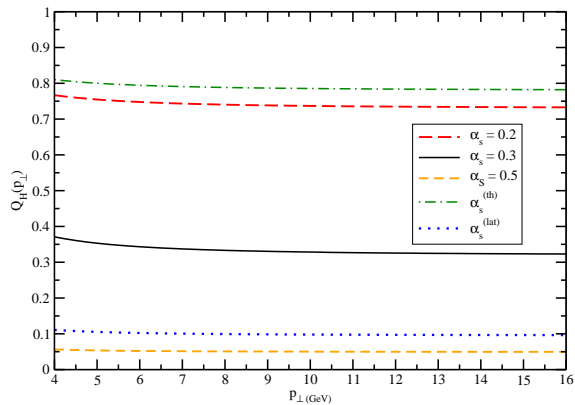}}
\vspace*{8pt} \caption{Quenching factor for heavy quark
$p_\perp$-spectrum.} \label{fig6}
\end{figure}

For heavy quarks we use the $p_\perp$ distribution of charmed
hadrons ($D$-mesons) produced in hadron collisions. It is
experimentally found \cite{alves} to be well described by the
following simple parametrization
\begin{equation}
\frac{dN^{\rm{vac}}_H}{ d^2p_\perp}\, = \, C \left (\frac{1}{bM_c^2+p^2_\perp}
\right )^{n/2} \, \, \, ,
\label{hparam}
\end{equation}
where $b=1.4\pm0.3$, $n=10.0\pm1.2$ and $M_c=1.5$ GeV. The quenching
factor for heavy quarks is shown in Fig. \ref{fig6}. The results are
similar to those obtained for light quarks in Fig. \ref{fig5}. While
the thermal QCD $\alpha_s$ prescription gives the higher quenching
factor, the $\alpha_s = 0.5$ one gives the smaller factor. The
lattice QCD prescription implies a small quenching factor $Q_H
\approx 0.1$, shown a strong suppression in the charm spectra when
this prescription is used. Again, at smaller values of $\alpha_s$,
higher the quenching factor due to elastic scattering in the QGP.

\begin{figure}[h]
\vspace*{0.5cm}
\centerline{\psfig{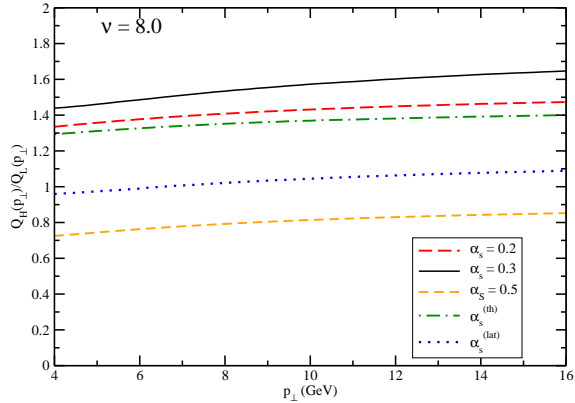}}
\vspace*{8pt} \caption{Heavy-to-light quenching factor ratio.}
\label{fig7}
\end{figure}

Recently, the ratio between the spectra of hadrons with heavy quarks
and with light quarks has been proposed as a tool to investigate the
medium formed in heavy ion collisions \cite{dokskhar}. Because of
its large mass, radiative energy loss for heavy quarks would be
lower than for light quarks. It occurs due to combined mass effects
\cite{dokskhar,wang2zhang}: the reduction of the formation time of
gluon radiation and  the  suppression of gluon radiation at angles
smaller than the ratio of the quark mass to its energy by
destructive quantum interference \cite{deadcone} - the dead-cone
effect.  The predicted consequence of these distinct radiative
energy losses is an enhancement of the heavy to light spectra ratio
at moderately large transverse momentum, relative to that observed
in the absence of energy loss (A recent analysis for LHC energies is
given in Ref. \cite{daineseEPJC}). As the behavior of this ratio
considering collisional energy loss is still an open question, in
Fig. \ref{fig7} we present the ratio between heavy and light quark
quenching factors, which reflects the heavy to light hadrons
$(D/\pi)$ ratio, considering only collisional energy loss. While for
values of fixed $\alpha_s$ smaller than 0.5, the results show an
enhancement factor close to $1.4$, this enhancement is suppressed
when running coupling prescriptions are considered. This feature
could suggest that with an expanding cooling medium, the collisional
energy loss for heavy and light quarks would be of similar
magnitude. Finally, if a large value of $\alpha_s$ is considered the
ratio can be smaller than one.

\section{Conclusions}
\label{conc}

Before we summarize,  let us discuss the  assumptions made in this
work in order to simplify the calculations, which may affect our
results. First, we have disregarded the momentum dependence of the
drag and diffusion coefficients, which contain the dynamics of the
elastic collisions, replacing it by its average value. Second, the
entire discussion is based on the one dimensional Fokker-Planck
equation and the Bjorken model for a nuclear collision, which may
not be a very realistic description, providing only a qualitative
estimate. An extension for three dimensional analysis is still in
discussion and may lead to a revision of our conclusions. Moreover,
the inclusion of diagrams for other channels in  elastic processes
implies an additional uncertainty. The perturbative expressions for
the (radiative and collisional) partonic energy loss, which are only
known to leading order in the strong coupling constant, may get
substantial corrections of higher order. A calculation of
next-to-leading order corrections to energy loss in perturbative QCD
would be desirable to test the stability of our results and
conclusions. Finally, in order to compare our predictions with the
experimental data the fragmentation of partons into hadrons should
be considered and included in our calculations.

As a summary, in this paper we have investigated  the dependence of
the parton collisional energy loss in a QGP on the value of the
strong coupling. Since the plasma is not a static medium, a fixed
value for $\alpha_s$ is a crude approximation. More realistic
estimates should take into account  the evolution of the fireball.
We have considered running coupling in the calculation, evolving it
with the cooling of the QGP. From the Fokker-Planck equation, we
derived the transport coefficients and related them with parton mean
energy loss. The drag coefficient is found to be modified with the
value of $\alpha_s$ considered, and it is strongly dependent on the
running coupling prescription used in the analysis. The lattice one
gives larger drag coefficient values, while the thermal QCD one
gives smaller value, always compared with the fixed value used in
previous calculations. A similar feature is verified in the mean
energy loss results and quenching factors. For light quarks, we
found that the radiative and collisional energy loss are of the same
order of magnitude, in the high $p_\perp$ region, if the $\alpha_s$
value is larger  than $0.3$. For smaller values of coupling, the
gluon bremsstrahlung becomes the dominant process for energy loss,
again. The ratio between heavy and light quenching factors has been
studied, and we have found  the absence of enhancement  if running
coupling is used. It was a striking result, since it suggest that
heavy and light quarks have the same order of magnitude for the
collisional energy loss. Our results motivate a similar study in
radiative parton energy loss.

\section*{Acknowledgments}

This work was partially financed by CNPq and FAPERGS, Brazil.



\begin{thebibliography}{00}

\bibitem{adams} J. Adams {\it et al.} [STAR Collaboration],   {\it Nucl.\ Phys.\ A}{\bf 757}, 102 (2005).

\bibitem{rafelski} J. Letessier and J. Rafelski, {\em Hadrons and
  Quark Gluon Plasma}, Cambridge Monogr. Part. Phys. Nucl. Phys.
  Cosmol. 18, 1 (2002).

\bibitem{wong} C. Y. Wong, {\em Introduction to High-Energy Heavy Ion
  Collisions}, Singapore, Singapore: World Scientific (1994).

\bibitem{wang_jacobs} P.~Jacobs and X.~N.~Wang,
  {\it Prog.\ Part.\ Nucl.\ Phys.}{\bf 54}, 443 (2005).

\bibitem{baier_schiff} R.~Baier, D.~Schiff and B.~G.~Zakharov,
    {\it Ann.\ Rev.\ Nucl.\ Part.\ Sci.}{\bf 50}, 37 (2000).

\bibitem{wang_QGP3} M.~Gyulassy, I.~Vitev, X.~N.~Wang and B.~W.~Zhang,
  arXiv:nucl-th/0302077.

\bibitem{kovner_QGP3} A.~Kovner and U.~A.~Wiedemann,
  arXiv:hep-ph/0304151.

\bibitem{hard_probes_cern} A.~Accardi {\it et al.},
  arXiv:hep-ph/0310274.

\bibitem{wang_prc_energy} X.~N.~Wang,
  {\it Phys.\ Rev.\ C}{\bf 70}, 031901 (2004).

\bibitem{gyulassy_adil} A.~Adil and M.~Gyulassy,
  {\it Phys.\ Lett.\ B}{\bf 602}, 52 (2004).

\bibitem{armesto_dainese} N.~Armesto, A.~Dainese, C.~A.~Salgado and U.~A.~Wiedemann,
  {\it Phys.\ Rev.\ D}{\bf 71}, 054027 (2005).

\bibitem{wang2zhang} B.~W.~Zhang, E.~Wang and X.~N.~Wang,
  {\it Phys.\ Rev.\ Lett.}{\bf 93}, 072301 (2004).

\bibitem{gyulassy_magda} M.~Djordjevic and M.~Gyulassy,
  {\it Nucl.\ Phys.\ A}{\bf 733}, 265 (2004).

\bibitem{wang_nonabelian} Q.~Wang and X.~N.~Wang,
  {\it Phys.\ Rev.\ C}{\bf 71}, 014903 (2005).

\bibitem{rapp} H.~van Hees and R.~Rapp,
  {\it Phys.\ Rev.\ C}{\bf 71}, 034907 (2005)

\bibitem{guy_moore} G.~D.~Moore and D.~Teaney,
  {\it Phys.\ Rev.\ C}{\bf 71},  064904 (2005).

\bibitem{mustafa_thoma} M.~G.~Mustafa and M.~H.~Thoma,
  {\it Acta Phys.\ Hung.\ A}{\bf 22},  93 (2005)
  [arXiv:hep-ph/0311168].

\bibitem{mustafa} M.~G.~Mustafa,
  {\it Phys.\ Rev.\ C}{\bf 72}, 014905 (2005).

\bibitem{stopping}A.~K.~Dutt-Mazumder, J.~e.~Alam, P.~Roy and B.~Sinha,
  {\it Phys.\ Rev.\ D}{\bf 71}, 094016 (2005).


\bibitem{vitev} I.~Vitev and M.~Gyulassy,
  {\it Phys.\ Rev.\ Lett.}{\bf 89}, 252301 (2002)

\bibitem{salgado} C.~A.~Salgado and U.~A.~Wiedemann,
  {\it Phys.\ Rev.\ Lett.}{\bf 89}, 092303 (2002)

\bibitem{wongAlpha} S.~M.~H.~Wong,
  {\it Phys.\ Rev.\ C}{\bf 56}, 1075 (1997).

\bibitem{nayak_mclerran} G.~C.~Nayak, A.~Dumitru, L.~D.~McLerran and W.~Greiner,
  {\it Nucl.\ Phys.\ A}{\bf 687},  457 (2001)

\bibitem{bjorken} J. D. Bjorken, {\it Phys. Rev. D}{\bf 27}, 140 (1983).

\bibitem{gunion_vogt}  J. F. Gunion, R. Vogt, {\it Nucl. Phys. B}{\bf 492}, 301 (1997) .

\bibitem{vogt_levai} P.~Levai and R.~Vogt,
  {\it Phys.\ Rev.\ C}{\bf 56}, 2707 (1997)

\bibitem{kaczamarek} O. Kaczmarek, F. Karsch, F. Zantow, P. Petreczky,
  {\it Phys. Rev. D}{\bf 70}, 074505 (2004) .

\bibitem{rafelski_tounsi} J.~Letessier, A.~Tounsi and J.~Rafelski,
  {\it Phys.\ Lett.\ B}{\bf 389}, 586 (1996).

\bibitem{wong_spectra} S.~M.~H.~Wong,
  {\it Phys.\ Rev.\ C}{\bf 58}, 2358 (1998).

\bibitem{denterria} D. d'Enterria, D. Peressounko,
  {\it Eur.\ Phys.\ J.\ C}{\bf 46}, 451 (2006).

\bibitem{svetitsky} B.~Svetitsky,
  {\it Phys.\ Rev.\ D}{\bf 37}, 2484 (1988) .

\bibitem{dipali} M. G. Mustafa, D. Pal and D. K. Srivastava,
  {\it Phys. Rev. C}{\bf 57},  889 (1998).

\bibitem{zakharov_jetp_lett}  B.~G.~Zakharov,
  {\it JETP Lett.}{\bf 80}, 67 (2004)
  [{\it Pisma Zh.\ Eksp.\ Teor.\ Fiz.}{\bf 80},  75 (2004)]

\bibitem{bjorken_1982}  J.~D.~Bjorken,
  FERMILAB-PUB-82-059-THY.

\bibitem{thoma_ela} M. H. Thoma, {\it Phys. Lett. B}{\bf 273},  128 (1991).

\bibitem{braaten_thoma} E.~Braaten and M.~H.~Thoma,
  {\it Phys.\ Rev.\ D}{\bf 44},  2625  (1991).

\bibitem{peshier}A. Peshier,  arXiv:hep-ph/0605294; arXiv:hep-ph/0607275;  arXiv:hep-ph/0607299.

\bibitem{bdms}  R. Baier, Y. L. Dokshitzer, A. H. Mueller, and D. Schiff,
  {\it JHEP}{\bf 0109},  033 (2001).

\bibitem{mueller} B. M\"uller, {\it Phys. Rev. C}{\bf 67}, 061901(R) (2003) .

\bibitem{mrom} S.~Mrowczynski,
  {\it Phys.\ Lett.\ B}{\bf 269},  383 (1991).

\bibitem{thoma_epjc}  M.~H.~Thoma,
  {\it Eur.\ Phys.\ J.\ C}{\bf 16},  513 (2000)

\bibitem{braaten_pisa}  E.~Braaten and R.~D.~Pisarski,
  {\it Phys.\ Rev.\ Lett.}{\bf 64},  1338 (1990);
  {\it Nucl.\ Phys.\ B}{\bf 337}, 569 (1990); {\it B}{\bf 339},  310 (1990).

\bibitem{rafelski_prl}  D.~B.~Walton and J.~Rafelski,
  {\it Phys.\ Rev.\ Lett.}{\bf 84}, 31 (2000)

\bibitem{alam_prl}  J.~Alam, B.~Sinha and S.~Raha,
  {\it Phys.\ Rev.\ Lett.}{\bf 73},  1895 (1994).

\bibitem{alam_npa}  P.~Roy, J.~e.~Alam, S.~Sarkar, B.~Sinha and S.~Raha,
  {\it Nucl.\ Phys.\ A}{\bf 624},  687 (1997)

\bibitem{bales} R. Balescu, {\it Equilibrium and Non-Equilibrium Statistical Mechanics} (Wiley, New York, 1975).

\bibitem{william} W. E. Williams, {\it Partial Differential Equations} (Clarendon Press, Oxford, 1980).

\bibitem{dokskhar} Y. L. Dokshitzer and D. E. Kharzeev, {\it Phys.\ Lett.\ B}{\bf 519},  199 (2001).

\bibitem{alves} C. A. Alves {\it et al.} [E796 Collaboration], {\it Phys. Rev. Lett.}{\bf 77},  2392 (1996).

\bibitem{deadcone} Yu. L. Dokshitzer, V. A. Khoze, S. I. Troyan, {\it J. Phys. G}{\bf 17}, 1602 (1991).

\bibitem{daineseEPJC} A. Dainese, {\it Eur. Phys. J. C}{\bf 33},  495 (2004).

\end{thebibliography}
\end{document}